\documentclass[structabstract]{aa}
\usepackage{graphics,graphicx}
\usepackage{amssymb}
\usepackage{rotating}
\usepackage{natbib}
\usepackage{savesym}
\usepackage{amsmath}
\savesymbol{iint}
\usepackage{txfonts}
\restoresymbol{TXF}{iint}

\bibpunct{(}{)}{;}{a}{}{,}

\begin{document}

\title{{\it XMM-Newton} view of the N~206 superbubble in the Large Magellanic Cloud}


\author{P. J. Kavanagh\inst{1}
  \and M. Sasaki\inst{1}
                         \and S. D. Points\inst{2}
                         }

\offprints{P. J. Kavanagh, \email{kavanagh@astro.uni-tuebingen.de}}

\institute{Institut f\"{u}r Astronomie und Astrophysik, Kepler Center for Astro and Particle Physics, Eberhard Karls Universit\"{a}t, 72076 T\"{u}bingen, Germany 
    \and Cerro Tololo Inter-American Observatory, Casilla 603, La Serena, Chile
}

\date{Received - / Accepted -}

\abstract{}{We perform an analysis of the X-ray superbubble in the N~206 \ion{H}{ii} region in the Large Magellanic Cloud using current generation facilities to gain a better understanding of the physical processes at work in the superbubble and to improve our knowledge of superbubble evolution.}{We used \textit{XMM-Newton} observations of the N~206 region to produce images and extract spectra of the superbubble diffuse emission. Morphological comparisons with H$\alpha$ images from the Magellanic Cloud Emission Line Survey were performed, and spectral analysis of the diffuse X-ray emission was carried out. We derived the physical properties of the hot gas in the superbubble based on the results of the spectral analysis. We also determined the total energy stored in the superbubble and compared this to the expected energy input from the stellar population to assess the superbubble growth rate discrepancy for N~206.}{We find that the brightest region of diffuse X-ray emission is confined by a H$\alpha$ shell, consistent with the superbubble model. In addition, faint emission extending beyond the H$\alpha$ shell was found, which we attribute to a blowout region. The spectral analysis of both emission regions points to a hot shocked gas as the likely origin of the emission. We determine the total energy stored in the bubble and the expected energy input by the stellar population. However, due to limited data on the stellar population, the input energy is poorly constrained and, consequently, no definitive indication of a growth rate discrepancy is seen.}{Using the high-sensitivity X-ray data from \textit{XMM-Newton} and optical data from the Magellanic Cloud Emission Line Survey has allowed us to better understand the physical properties of the N~206 superbubble and address some key questions of superbubble evolution.}

\titlerunning{{\it XMM-Newton} view of the N~206 superbubble}
\maketitle 

\section{Introduction}
Massive stars, via their fast stellar winds and subsequent supernova explosions, are responsible for energising and enriching the interstellar medium (ISM), and are the source of the ISM's dynamic hot-phase. Given that massive stars usually form in groups, their collective mechanical output into the surrounding ISM creates so-called `superbubbles', 100-1000 pc diameter shells of swept-up interstellar material that contains a hot (10$^{6}$ K), shock-heated X-ray emitting gas. Superbubbles are one of the primary engines for driving the morphology and evolution of the multi-phase ISM and are of key importance in the understanding of matter recycling in galaxies. \citet{maclow1988} showed that these objects could be considered scaled-up versions of stellar wind blown bubbles and expanded on the classical analytical model of \citet{Weaver1977} in this area. More recent numerical models of superbubbles are capable of predicting temperature distributions and X-ray luminosities \citep[see][for example]{Breit2006,Rod2011}.\\

\par Owing to the soft nature of superbubble X-rays ($<$ 2 keV) and the location of massive star forming regions in the Galactic disk, these objects in our Galaxy are almost completely obscured by foreground absorbing material. Conversely, the superbubble population of the \object{Large Magellanic Cloud} (LMC) is ideal for study. The \object{LMC}, which is a dwarf irregular galaxy with indications for spiral structures, is one of the closest neighbours to our Galaxy. The distance to the LMC of 48 kpc \citep{Macri2006}, its modest extinction in the line of sight (average Galactic foreground $N_{\rm{H}} \approx 7 \times 10^{20}\ \rm{cm}^{-2}$), and its small inclination angle of $\sim 40^{\circ}$ \citep{Feast1991} make it an ideal laboratory for the multiwavelength study of superbubbles. This allows one to trace the diffuse X-ray emitting gas and constrain the mass and energy input to the superbubbles using the observed stellar populations. \\

\par \object{N~206} \citep[][also known as \object{DEM L221}]{Henize1956} is an \ion{H}{ii} region located in the southeast of the LMC that is excited by the winds of the massive stars in the young \object{NGC 2018} cluster, and the \object{LH 66} and \object{LH 69 OB} associations \citep{Lucke1970}. \citet{Gorjian2004} used {\it Spitzer} observations of N~206 to identify regions of very active star formation in the shell of cold material surrounding the \ion{H}{ii} region. More recently, \citet{Romita2010} used a multiwavelength approach to identify and classify many of the young stellar objects (YSOs) in N~206 and determined an above average star formation rate of some $\sim 5$ times higher than the LMC as a whole. This \ion{H}{ii} complex also harbours the supernova remnant SNR B0532-71.0 \citep{Math1973,Williams2005} and contains an X-ray superbubble. The diffuse X-ray emission in the \ion{H}{ii} region was first identified and attributed to a superbubble by \citet{Dunne2001} using \textit{ROSAT} HRI observations and H$\alpha$ images from PDS scans of the Curtis Schmidt plates of \citet{Kennicutt1986}. These authors found the X-ray emission to be confined by an H$\alpha$ structure, indicative of a superbubble scenario. \\

\par  In this paper we present an analysis of the N~206 superbubble using current generation facilities to gain a better understanding of the physical processes at work in the N~206 superbubble. In Sections 2 we outline the {\it XMM-Newton} and optical observations.  In Section 3 we descibe the data analysis including a morphological comparison to H$\alpha$ emission and spectral analysis of the superbubble X-ray emission. In Section 4 we discuss the results of the analysis and address some key evolutionary questions of the N~206 superbubble. A summary of the work is presented in Section 5.

 \begin{figure}
\begin{center}
\resizebox{\hsize}{!}{\includegraphics{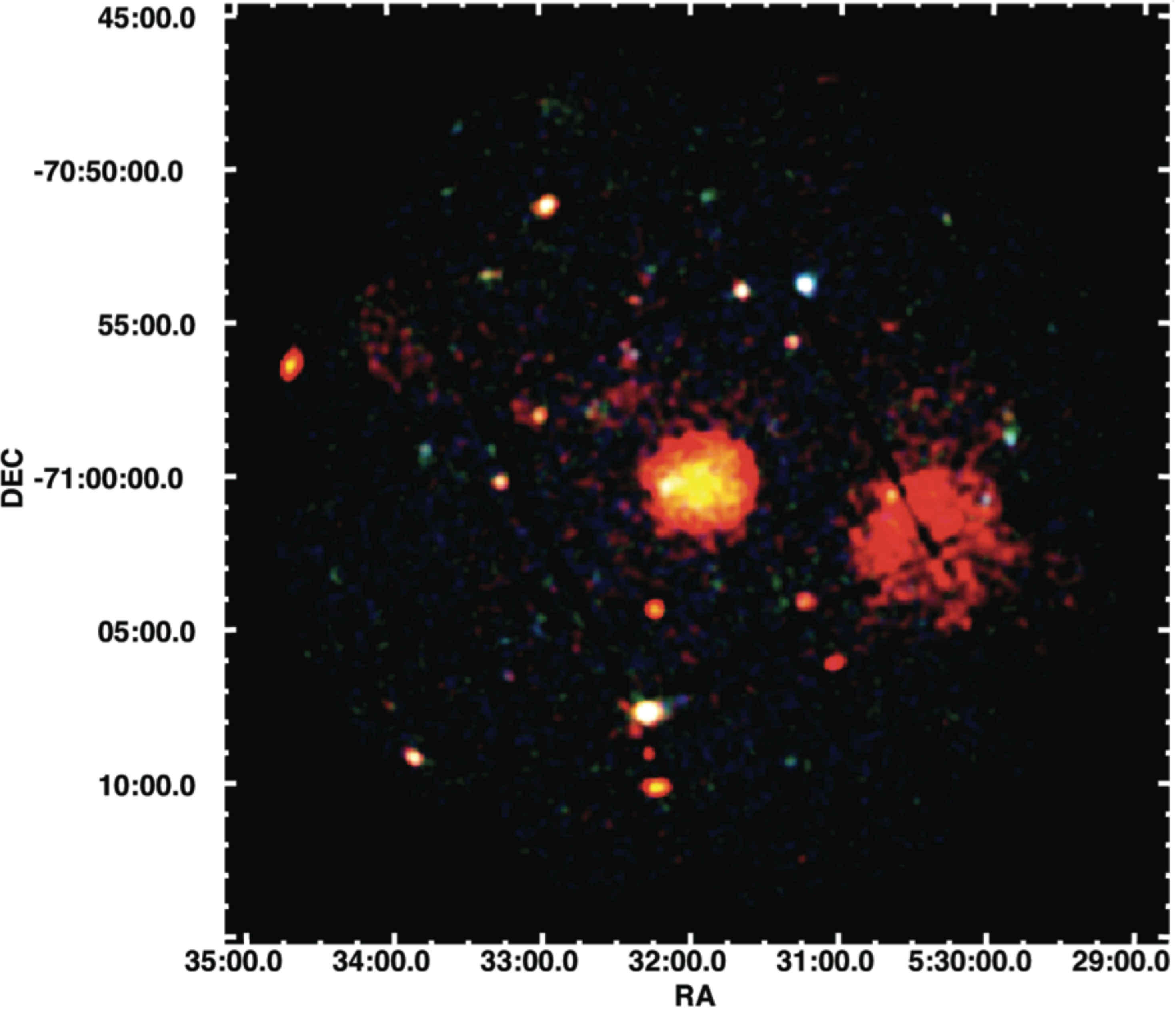}}
\caption{{\it XMM-Newton} EPIC false-colour mosaic image of SNR B0532-71.0 and the N~206 superbubble. Red corresponds to the 0.3-1 keV energy range, green to 1-2 keV, and blue to 2-8 keV. The images have been Gaussian smoothed with a 3-pixel kernel.}
\label{image}
\end{center}
\end{figure}

\section{Observations}
\subsection{X-rays}
Owing to the presence of SNR B0532-71.0 in the N~206 \ion{H}{ii} complex, the region has been observed with both \textit{Chandra} \citep{Weisskopf2002} and \textit{XMM-Newton} \citep{Jansen2001} with the SNR at the aimpoint. Unfortunately, the \textit{Chandra} data are not suitable for the purposes of our study because they do not cover the superbubble emission region. The European Photon Imaging Cameras \citep[EPIC,][]{Struder2001,Turner2001} onboard {\it XMM-Newton}  have observed SNR B0532-71.0 on two occasions (Obs. IDs 0089210101 and 0089210901, PI R. Williams). Because the PN camera operated in large-window mode for Obs. ID 0089210101, only a fraction of the N~206 superbubble was covered, making these data unsuitable for our analysis. The PN was not in operation during Obs. ID 0089210901. Hence, only MOS data are available for each observation which were in imaging full-frame mode for a total of $\sim51$ ks during the observations. We reduced each of the datasets using the standard reduction tasks in SAS 10.0.0, filtering for periods of high particle background, which resulted in a combined effective exposure of 32 ks (9 ks + 23 ks). We created mosaic images in the 0.3-1 keV, 1-2 keV, and 2-8 keV bands by combining the data from each MOS camera in each observation to produce a false-colour X-ray image of the region (shown in Fig. \ref{image}), which was smoothed using a Gaussian filter. The superbubble X-ray emission is clearly visible to the west of SNR B0532-71.0 in the centre of the field of view (FOV). The X-ray emission is very soft with the vast majority of the emission being $<$ 1 keV.

\subsection{Optical}
The Magellanic Cloud Emission Line Survey (MCELS) observations \citep{Smith2006} were made with the 0.6 m University of Michigan/Cerro Tololo Inter-American Observatory (CTIO) Curtis Schmidt Telescope equipped with a SITE 2048 $\times$ 2048 CCD, producing individual images of $1.35^{\circ} \times 1.35^{\circ}$ at a scale of 2.3$\arcsec$ pixel$^{-1}$. The survey mapped both the LMC ($8^{\circ} \times 8^{\circ}$) and the SMC ($3.5^{\circ} \times 4.5^{\circ}$) in narrow bands covering [\ion{O}{iii}]$\lambda$5007 \AA, H$\alpha$, and [\ion{S}{ii}]$\lambda$6716,6731 \AA in addition to matched green and red continuum bands, which are used primarily for the subtraction of the stellar continuum from the narrow band images. The survey data were flux-calibrated and combined to produce mosaicked images. Cutouts from the mosaics of each LMC emission line image around N~206 were used for our analysis.

 \section{Data analysis}
 \subsection{X-ray point sources}
\begin{figure}
\begin{center}
\resizebox{\hsize}{!}{\includegraphics{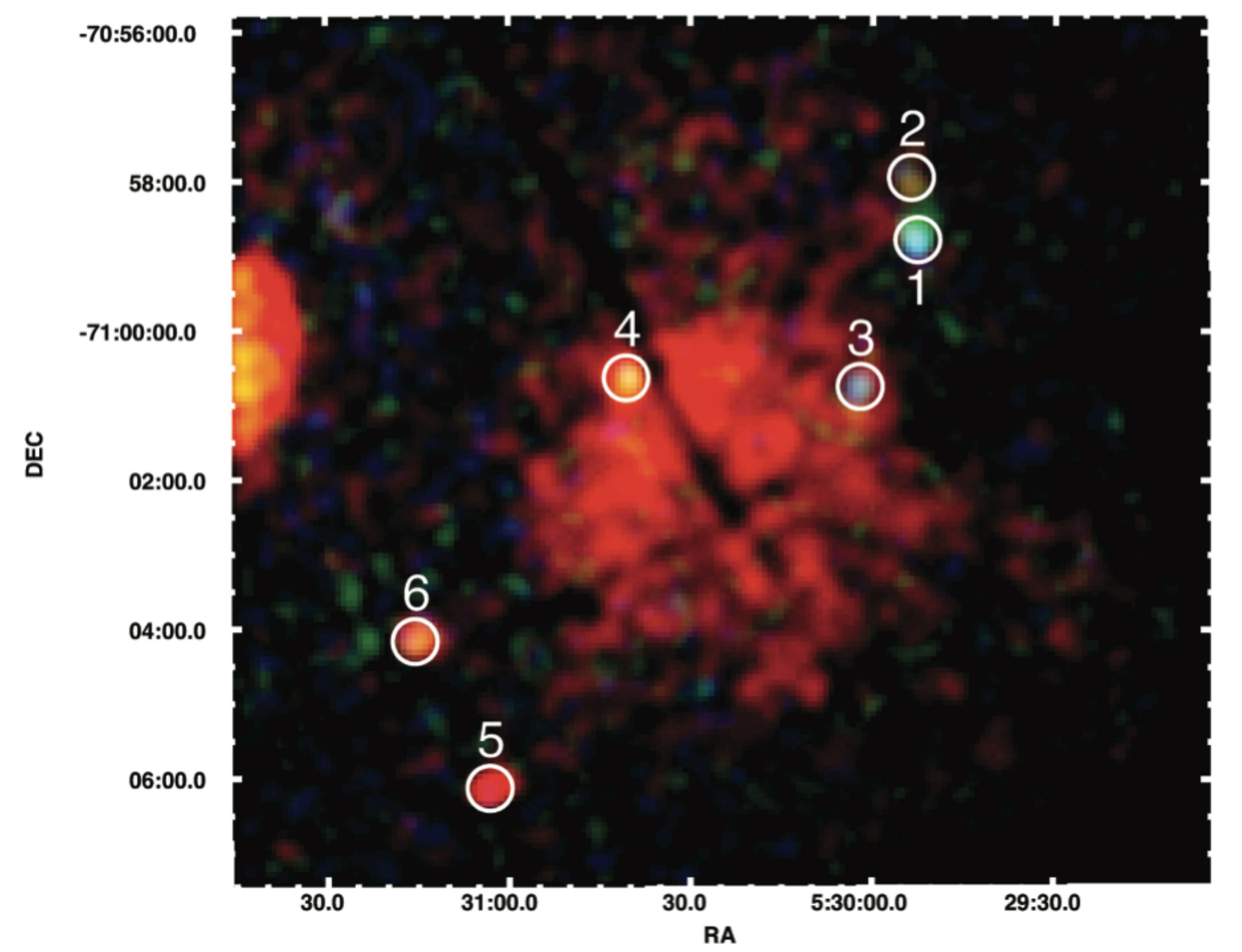}}
\caption{Detected X-ray sources plotted on a cutout from the false-colour X-ray mosaic image in Fig. \ref{image} in the region of the N~206 superbubble. Red corresponds to the 0.3-1 keV energy range, green to 1-2 keV, and blue to 2-8 keV. See Table \ref{pointsources} for source details.}
\label{psources}
\end{center}
\end{figure}

Source detection was performed using the SAS detection metatask {\tt edetect\_chain} across various energy bands to improve detection sensitivity. Several sources were detected in the region of the superbubble, which are shown in Fig. \ref{psources} with their multiwavelength counterparts obtained from the SIMBAD database given in Table \ref{pointsources}. We detected a high-mass X-ray binary (HXMB) \object{XMMU J052952.9-705850} (Source 1) and a candidate HMXB \object{USNO-B1.0 0189-00079195} (Source 3), both of which were identified by \citet{Shty2005}. Each of these sources is likely associated with the N~206 \ion{H}{ii} complex. Additionally, we detected an X-ray source in the NGC 2018 cluster, namely the O4-5III(f) star \object{HD 269676}. The remaining sources include the foreground star \object{HD 269669}, the likely background AGN \object{[SHP2000] LMC 211} \citep{Sasaki2000}, and a source without a counterpart. Detected sources located in the regions of diffuse emission were masked in the subsequent spectral analysis based on the optimum extraction regions determined using the SAS task {\tt eregionanalyse}.

\begin{table*}
\caption{X-ray point sources in the superbubble region with SIMBAD counterparts}
\begin{center}
\begin{small}
\label{pointsources}
\begin{tabular}{lllll}
\hline
Source &	RA & DEC &  SIMBAD counterpart & Type \\
\hline
\hline
1 &	05:29:52.800 & -70:58:46.56 &	 XMMU J052952.9-705850	&	HMXB \\

2 &	05:29:54.192 & -70:58:04.08 & 	-	&	 - \\

3 &	05:30:02.016 & -71:00:46.80 &		USNO-B1.0 0189-00079195	&	cand. HMXB \\

4 &	05:30:40.632 & -71:00:39.24 &		[SHP2000] LMC 211	 &	X-ray source \\
	
5 &	05:31:02.976 & -71:06:09.00 &	HD 269669	&		G0 star \\

6 &	05:31:15.144 & -71:04:10.20 &	HD 269676	&		O4-5III(f) star \\
\hline
\end{tabular}
\end{small}
\end{center}
\end{table*}%

\subsection{Morphology}
\label{morph}
\begin{figure*}
\begin{center}
\resizebox{\hsize}{!}{\includegraphics{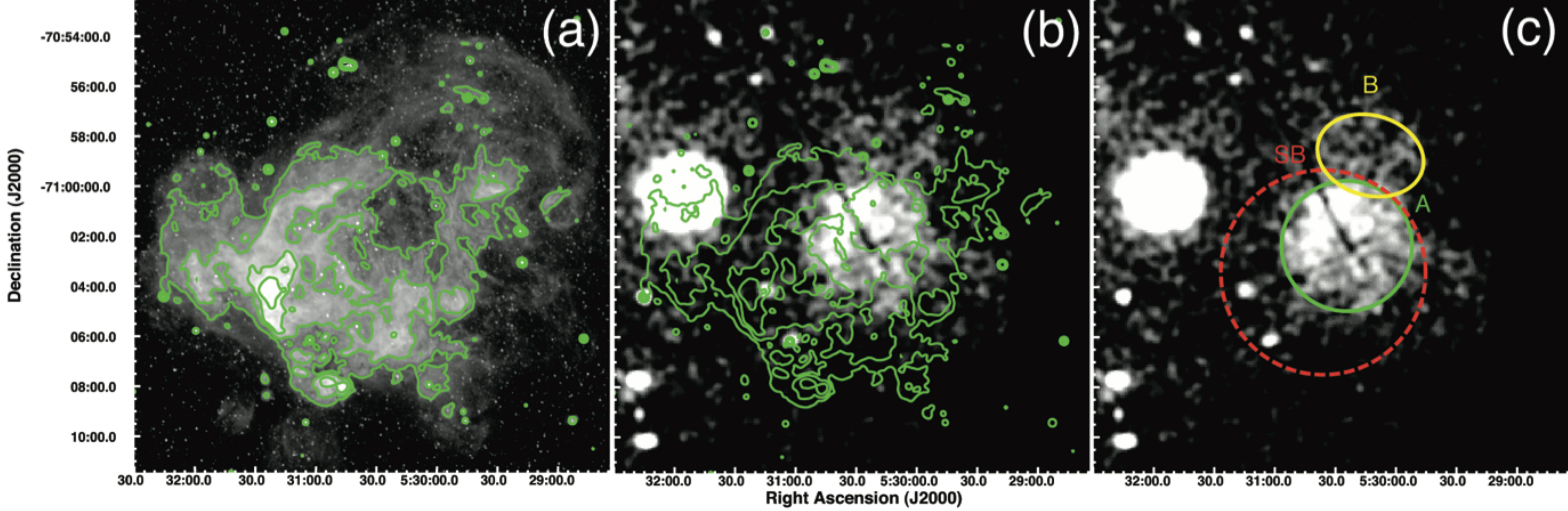}}
\caption{(a) - MCELS H$\alpha$ image with contours. The lower contour limit has been adjusted to highlight the H$\alpha$ shell that confines the X-ray emitting gas. The levels correspond to four logarithmically spaced intervals between this adjusted lower limit and the maximum pixel value. (b) - 0.3-1 keV \textit{XMM-Newton} image with the H$\alpha$ contours overlaid. - (c) 0.3-1 keV \textit{XMM-Newton} image with superbubble emission regions indicated.}
\label{multi}
\end{center}
\end{figure*}

We compared the soft band 0.3-1 keV X-ray image to the MCELS H$\alpha$ image of the superbubble to assess the superbubble morphologies, shown in Fig. \ref{multi}. We found that the brightest X-ray emitting region of the superbubble is confined by an H$\alpha$ structure, as found previously by \citet{Dunne2001}, labelled as Region A in Fig. \ref{multi}c. Additionally, we identified an apparent blowout region to the north (labelled as Region B in Fig. \ref{multi}c) that is surrounded by a fainter and more dispersed H$\alpha$ structure. The presence of a blowout is also supported by the IR images of \citet{Gorjian2004}, which show that the cooler material surrounding the H$\alpha$ shell is also dispersed in the region of the blowout. The H$\alpha$ structure confining Region A appears to be part of a large approximately circular H$\alpha$ shell (labelled as Region SB in Fig. \ref{multi}c) with $\sim8$\arcmin diameter (corresponding to $\sim112$ pc at the LMC distance), which encompasses the OB associations in the \ion{H}{ii} region. Interestingly, no X-ray emission is detected from the majority of Region SB. This is somewhat unusual, as we would not expect the hot X-ray emitting gas to be distributed with such non-uniformity within the superbubble. Likely reasons for this are discussed in Section \ref{emission}.

\subsection{EPIC spectra}
Spectral analysis of extended diffuse emission is complicated by the contributions of both the instrumental and astrophysical backgrounds. In the case of the astrophysical background, the observational FOV of our analysis is largely free of extended diffuse emission apart from the superbubble. Hence, a background extracted from these data should adequately account for the astrophysical background, given that we can assume it to be uniform over such a small area. The only problem with such a background selection in this case is the variation in spectral response of the detector, which can be easily corrected for. However, this may not be adequate to account for the instrumental background. The instrumental background below 2 keV for the EPIC instruments is dominated by fluorescence lines caused by the interaction of high-energy particles with the material surrounding the detectors  \citep{Lumb2002,Read2003}. For the MOS cameras these are the Al K$\alpha$ and Si K$\alpha$ lines at $\sim1.49$ keV and $\sim1.75$ keV, respectively. Variation in the strength of these lines across the detector makes the background region subject to a different instrumental background than the superbubble. One background selection method that can account for this effect is the use of the so-called `blank-sky' data. The {\it XMM-Newton} EPIC background working group has produced these blank-sky data sets for each EPIC instrument over all the read-out modes and filters \citep{Carter2007}. The blank-sky data have been produced by merging several observations and removing point sources to leave an observational dataset representative of the average astrophysical background and the instrumental background. Thus, using these blank-sky data, we can select a background from the same region on the detector as the superbubble in our observational dataset, which can adequately account for the instrumental background and the astrophysical background, assuming the averaged astrophysical background from the blank-sky data is representative of the astrophysical background in our observations.\footnote{A detailed comparison of the different methods to estimate the backgrounds can be found in \citet{Sasaki2004}.}

\par We assessed the effect of each of these background selection techniques on the results of spectral fits. We found that, apart from small residuals at the fluorescence emission line energies, the background-subtracted spectra in each case were almost identical with the results of the spectral fits, being consistent to within the confidence range of the fit parameters. Hence, given that the blank-sky data more accurately accounted for the instrumental background, we discuss the results obtained using this method. 

\subsubsection{Spectral fits}
\label{fits-section}
\par  We extracted X-ray spectra from Region A and Region B  (with point sources masked). Spectral analysis was performed using Xspec version 12.5.1\footnote{Note: Throughout the analysis spectra from both the MOS cameras for each observation were fitted simultaneously in Xspec.} \citep{Arnaud1996}. Initial fits on the regions showed that Region B is adequately fitted with either a soft collisional ionisation equilibrium model \citep[CIE, the APEC model in Xspec,][]{Smith2001}, or a soft non-equilibrium ionisation model \citep[NEI, the NEI model in Xspec,][]{Bork2000}, whereas Region A can be fitted with either a moderately hard single NEI model, a moderately hard NEI + soft CIE model, or a two NEI model, with the two-component models yielding the better fit statistics. The single-component model parameters of Region B were very similar to the softer component in the two-component fit to the Region A spectra. Therefore we conclude that the soft components in Regions A and B are due to the same diffuse source present in each region. Because of this and because of the faint, dispersed H$\alpha$ structure surrounding Region B, we suspect that the soft X-ray emission in Regions A and B is due to a blowout region where hot gas is leaking from the superbubble interior into the surrounding regions. Given that the emission from the superbubble itself is best described by an NEI model, we favour this for describing the gas that escapes from the bubble and thus only the two component NEI model is considered for the remainder of the analysis.\\

\par We refitted the spectra linking the softer NEI model parameters in Regions A and B to obtain better statistics on the soft component, while the harder NEI component was used to fit Region A only. The normalisations of the linked soft NEI components were adjusted to account for the difference in volume of the emitting regions. Because we assumed that the soft NEI emission comes from the hot gas in the blowout, the plasma density is uniform across both regions, only varying in emitting volume. Given that the model normalisation is defined as

\begin{equation}
\label{norm}
norm = \dfrac{1}{10^{14}\times4\pi D^{2}}\int n_{e}n\ f_{gas}\ dV,
\end{equation}

\noindent where $f_{gas}$ is the filling parameter of the X-ray emitting gas, the ratio of the normalisations is that of the volumes. We assumed that the volume of the entire blowout (i.e., across Regions A and B) is defined by an ellipsoid $4\arcmin$ in height and $8\arcmin$ in width, determined from our X-ray images. Since only two-dimensional information is available for the blowout, we must assume its depth, which was set equal to its height of 4\arcmin. Thus, we have ellipsoidal radii of $a = c = 2\arcmin$ and $b = 4\arcmin$, or $a = c \approx 28$ pc and $b \approx 56$ pc at the distance of the LMC. To split the volume we cut the $b$ axis of the ellipsoid into $5\arcmin$ and $3\arcmin$ divisions, corresponding to the blowout volumes in Regions A and B, respectively. This is not a perfect solution, but without additional information the true dimensions of the blowout are impossible to determine.\\

\begin{figure}[ht]
\begin{center}
\resizebox{\hsize}{!}{\includegraphics[trim= 1.5cm 2cm 1.5cm 1.5cm, clip=true]{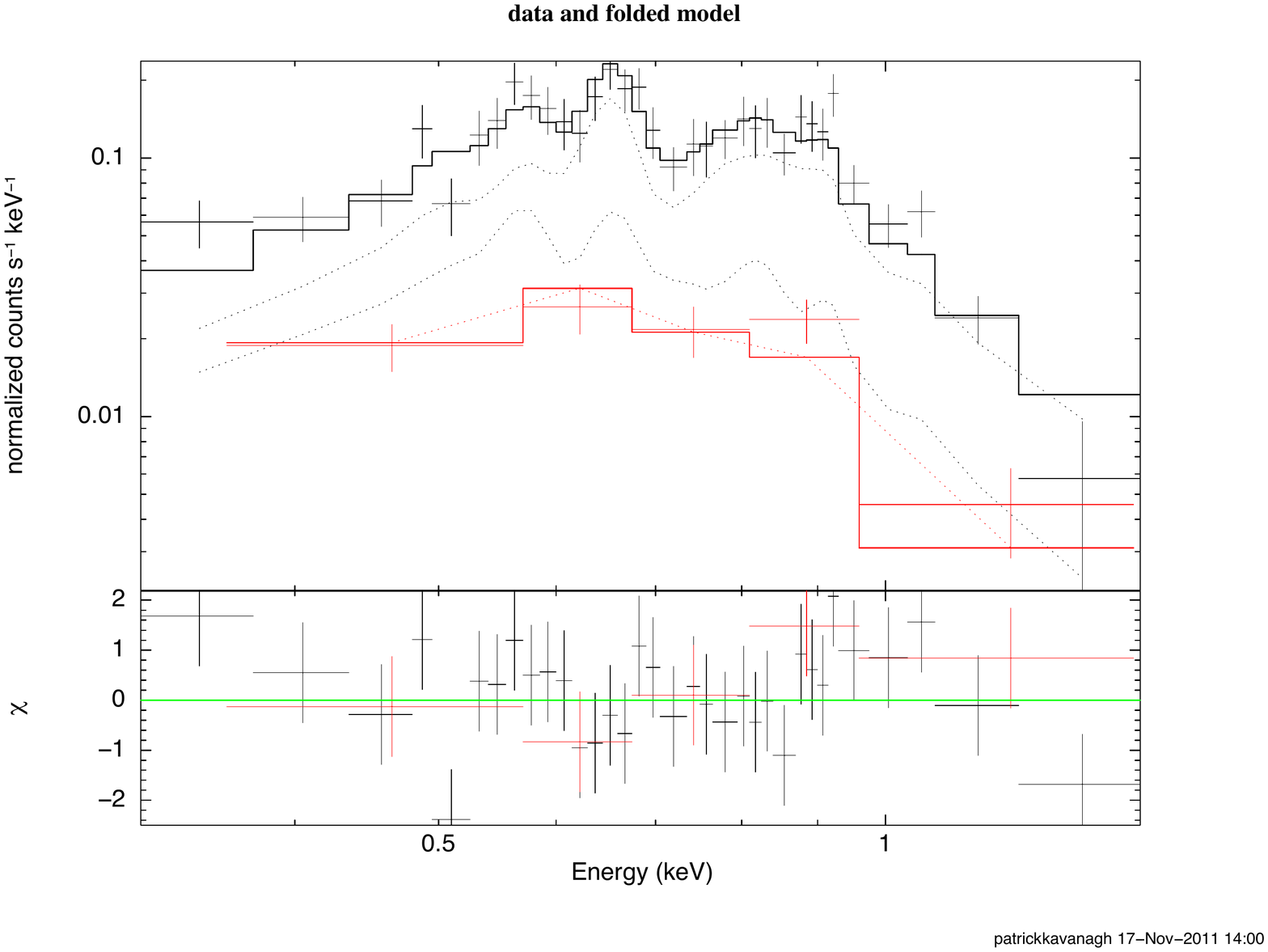}}
\caption{Best-fit 2 NEI model plotted on the MOS 2 spectra of the superbubble and blowout regions from the effective 23 ks observation. The spectrum of Region A is plotted in black, the spectrum of Region B in red. The components of the spectral fits are indicated by the dotted lines. The spectra are binned to a signal-to-noise ratio of 3.} 
\label{spectrum}
\end{center}
\end{figure}

\par The best-fit 2 NEI model is shown in Fig. \ref{spectrum}.  For clarity only the MOS 2 spectra from the effective 23 ks observation are shown. For completeness, the fit results for each of the models discussed are shown in Table \ref{fits}.  The absorbing hydrogen column ($N_{\rm{H}}$ ) for the fits was fixed at $0.7 \times 10^{21}$ cm$^{-2}$ \citep{Dickey1990} and the abundance was fixed at 0.5 solar. We chose to fix the $N_{\rm{H}}$ value for two reasons. Firstly, allowing the $N_{\rm{H}}$ parameter to vary results in a best-fit value lower than $0.7 \times 10^{21}$ cm$^{-2}$ but consistent within the upper 90\% confidence interval with the \citet{Dickey1990} value. Secondly, one could set the $N_{\rm{H}}$ value to that of the background AGN [SHP2000] LMC 211 (see Section \ref{emission}). However, with a derived value of $\sim (1 \pm 2) \times 10^{21}$ cm$^{-2}$, the $N_{\rm{H}}$ value for [SHP2000] 211, while consistent with the \citet{Dickey1990} value, is poorly constrained due to poor statistics. For these reasons we preferred to fix $N_{\rm{H}}$  at the known Galactic foreground value. \\

\begin{table*}[htdp]
\caption{Results of spectral fits}
\begin{center}
\begin{small}
\label{fits}
\begin{tabular}{lcccccccc}

\hline
 &  \multicolumn{3}{c}{CIE / NEI} &  \multicolumn{3}{c}{NEI} & \\
 &  \multicolumn{3}{c}{\hrulefill} &  \multicolumn{3}{c}{\hrulefill} & \\
Region (Model)  & $kT_{1}$ & $\tau_{1}$ & $norm_{1}$ & $kT_{2}$ & $\tau_{2}$ & $norm_{2}$ & $\chi^{2}/\nu$ \\

 &  (keV) &  ($10^{10}$ s cm$^{-3}$) & ($10^{-4}$) & (keV) &  ($10^{10}$ s cm$^{-3}$) &($10^{-4}$)  & \\ 
\hline
\hline
 &  & & & & & &   \\
A (NEI) & - & - & - & $1.01^{1.40}_{0.82}$ & $1.73^{2.26}_{1.27}$ & $1.09^{1.25}_{0.93}$  & 219.11/171=1.28 \\
 &  & & & & & &   \\
B (CIE) & $0.27^{0.29}_{0.25}$ & - & $1.57^{1.75}_{1.38}$ & - & - & -  & 46.20/56=0.83 \\
B (NEI) & $0.79^{2.02}_{0.47}$ & $2.19^{6.03}_{0.99}$ & $0.25^{0.41}_{0.16}$ & - & - & -  & 45.64/55=0.83 \\
 &  & & & & & &   \\
A+B (CIE+NEI) & $0.25^{0.27}_{0.24}$ & - & $2.73^{3.02}_{2.43}$ & $1.43^{2.08}_{0.99}$ & $1.52^{2.19}_{0.91}$ & $0.67^{0.83}_{0.59}$  & 265.48/227=1.17 \\
A+B (NEI+NEI) & $0.70^{1.09}_{0.48}$ & $1.74^{3.64}_{1.01}$ & $0.67^{0.99}_{0.58}$ & $0.99^{1.23}_{0.75}$ & $2.28^{3.14}_{1.46}$ & $1.29^{1.59}_{1.04}$  & 262.18/226=1.16 \\
\hline
\multicolumn{8}{l}{The upper and lower limits correspond to the 90\% confidence intervals of the fit parameters.}
\end{tabular}
\end{small}
\end{center}
\end{table*}%

\section{Discussion}
\subsection{Superbubble X-ray emission} 
\label{emission}
The morphological comparison between the X-ray and H$\alpha$ images in Fig. \ref{multi} shows the brightest region of X-ray emission confined by a H$\alpha$ shell structure, consistent with the superbubble picture of a hot diffuse gas confined by a cooler shell. However, in a rich stellar population such as that in the N~206 complex, an unresolved low-mass stellar population can account for a significant  fraction of the seemingly diffuse emission. Indeed, \citet{Oskinova2005} found that very early in the lifetime of a massive stellar population, the unresolved low-mass stellar population is the dominant contributor to the diffuse emission in the cluster.  However, the stellar populations in the N~206 superbubble are at a stage of evolution (see Section \ref{content}) where the truly diffuse emission is enhanced due to the higher mass loss rates of the post-main sequence population and SNRs. At this stage, the truly diffuse emission will dominate the apparent diffuse emission caused by the unresolved sources \citep{Oskinova2005}. To obtain an estimate for the contribution of unresolved low-mass population, we compared the N~206 stellar population to the well-known and characterised stellar population of the Orion Nebular Cluster (ONC), determined in the {\it Chandra} Orion Ultradeep Project \citep[COUP,][]{Getman2005}, a method which has been used for a multitude of stellar population analyses \citep[see][for example]{Getman2006,Ezoe2006,Broos2007,Kavanagh2011}. We assumed that the N~206 clusters and ONC have the same initial mass function (IMF), and X-ray luminosity distribution, differing only in the size of their underlying populations. This allowed us to scale the integrated X-ray luminosities of the lower mass bins of the ONC population to that of the N~206 population using the ONC mass stratified star counts and X-ray luminosities of \citet{Feigelson2005}. Additionally, since the X-ray luminosity of young low-mass stars can decrease by as much as $L_{\rm{X}} \propto \tau^{-0.75}$ \citep{Preibisch2005}, the ONC X-ray luminosity distribution was corrected for the increased ages of the N~206 populations (determined in Section \ref{content}). We chose the lowest age limit for the N~206 populations for this correction, which corresponds to the period of highest X-ray luminosity for the low-mass stars. The high-mass stellar populations of the N~206 clusters was used to extrapolate the assumed IMF to lower masses. The X-ray luminosity of the ONC population was then scaled based on the number of stars in the 1-3 M$_{\sun}$ mass bin. This provided us with an upper limit for the contribution of the unresolved low-mass population of $< 4 \times 10^{34}$ erg s$^{-1}$. Based on the spectral analysis in Section \ref{fits-section}, we determined the X-ray luminosity of the N~206 superbubble to be $\sim 7 \times 10^{35}$ erg s$^{-1}$. Hence, the unresolved low-mass population can at most contribute $\sim 6$ \% to the observed X-ray emission. Considering that this upper limit is already quite low, we ignore the contribution of the low-mass stars for the remainder of the analysis. \\

\par As discussed in Section \ref{morph}, no X-ray emission is detected from the majority of Region SB, which is unusual because we would expect the hot X-ray emitting gas to fill the entire superbubble volume. The likely explanation for this becomes apparent if we consider the spectra of the brightest point sources in Region A and in Region SB excluding Region A (which we will call Region SB$-$A), namely sources 4 and 6 identified as [SHP2000] LMC 211 and HD 269676, respectively. [SHP2000] LMC 211 is likely a background AGN, whereas HD 269676 is a member of the NGC 2018 cluster in the superbubble. We performed a spectral analysis of these sources and found [SHP2000] LMC 211 to have a significantly lower absorbing hydrogen column ($N_{\rm{H}} \sim (1 \pm 2) \times 10^{21}$ cm$^{-2}$) than HD 269676, whose corresponding value is several times higher ($N_{\rm{H}} \sim (9 \pm 4) \times 10^{21}$ cm$^{-2}$). This indicates that the line-of-sight absorption to the Region SB$-$A is likely sufficiently high to absorb the soft superbubble X-ray emission. We cannot exclude the possibility that some of the line-of-sight absorption to HD 269676 may be caused by stellar wind material in its circumstellar environment and, thus, the $N_{\rm{H}}$ value of $\sim (9 \pm 4) \times 10^{21}$ cm$^{-2}$ is an upper limit for the absorbing hydrogen column to Region SB-A. We note that for Galactic stars of similar spectral type this circumstellar absorption can be as high as $\sim5 \times 10^{21}$ cm$^{-2}$ \citep[see][for example]{Naze2011}. To estimate the effect that this additional absorption has on the superbubble X-ray emission from Region SB$-$A, we used the best-fit model of the detected X-ray emission from Region A (see Section \ref{fits-section}), added the additional absorption and considered the effect on the background-subtracted surface brightness of the emission. We find that if we assume no circumstellar absorption around HD 269676 (i.e., the superbubble X-ray emission is subject to the best-fit $N_{\rm{H}}$), the surface brightness in the 0.3-1 keV range is expected to be almost an order of magnitude lower than the detected superbubble emission. If we consider that circumstellar absorption contributes $\sim5 \times 10^{21}$ cm$^{-2}$, as is possible for Galactic stars of similar spectral type, then the surface brightness is still some 80\% lower than that of the detected superbubble emission, and is lower than half of the blowout emission in Region B, which itself is quite faint. \\

\par
\begin{figure}[ht]
\begin{center}
\resizebox{\hsize}{!}{\includegraphics{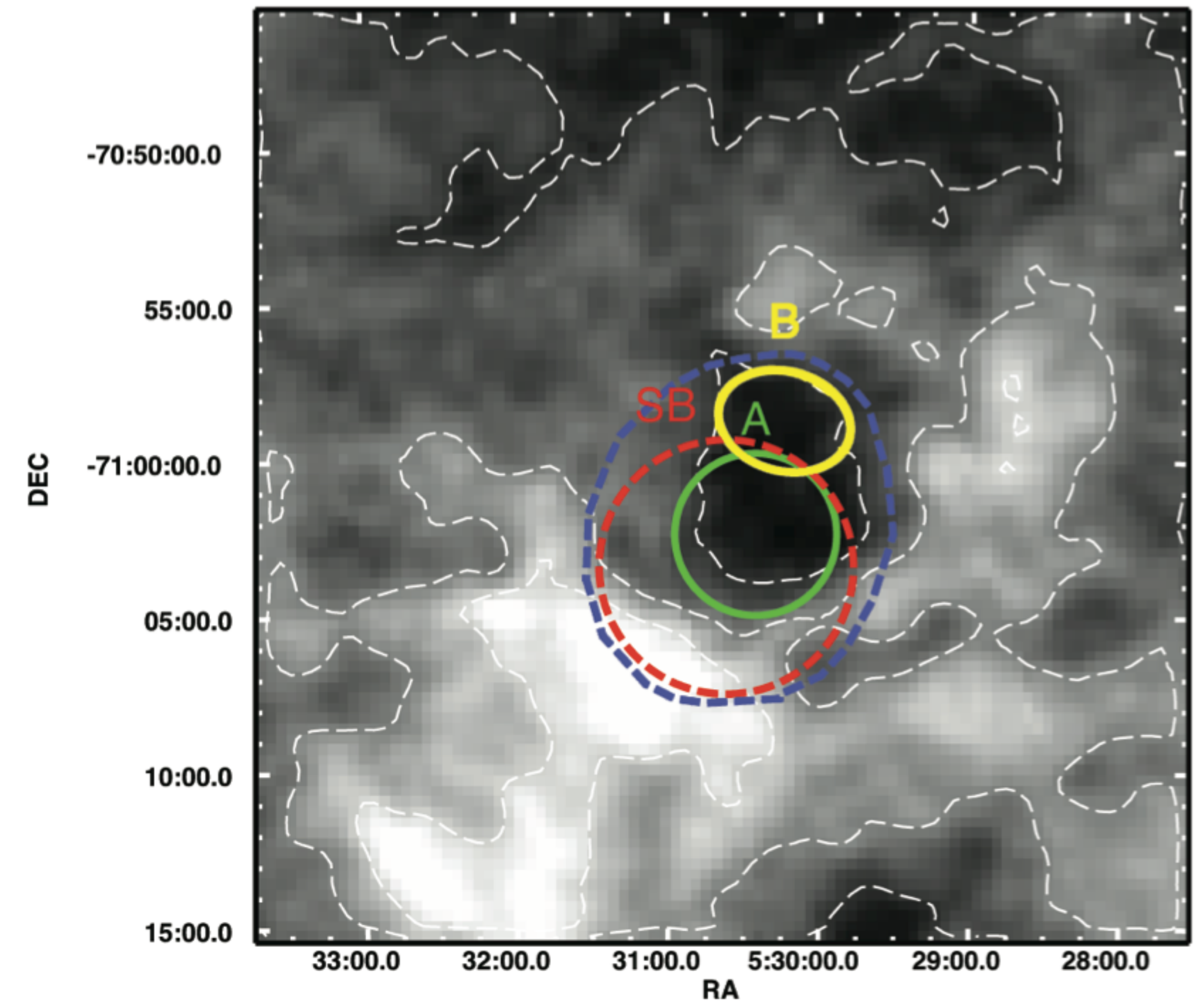}}
\caption{ATCA-Parkes 21 cm line map in the direction of N~206 from the \ion{H}{i} Magellanic Cloud Survey. The superbubble emission regions (A and B) and superbubble boundaries (SB) are indicated. The contours correspond to 25\%, 50\%, and 75\% of the peak emission. An additional region, shown here in blue, corresponds to the region from which we extracted the 21 cm line flux density for the analysis discussed in Section \ref{HIgas}.}
\label{HI_image}
\end{center}
\end{figure}

 Additional evidence of significant variation in line-of-sight absorbing material comes from the \ion{H}{i} 21 cm line map towards N~206. We obtained the ATCA-Parkes 21 cm emission image in the direction of N~206 from the \ion{H}{i} Magellanic Cloud Survey\footnote{Available at http://www.atnf.csiro.au/research/HI/mc/}. The ATCA and Parkes observations are outlined in \citet{Kim1998} and \citet{Kim2003}, respectively. The \ion{H}{i} 21 cm line map (shown in Fig. \ref{HI_image}) clearly illustrates the variation in absorbing material across N~206. The \ion{H}{i} gas is morphologically quite similar to the distribution of the 24 $\mu$m emitting dust in N~206 \citep{Gorjian2004} with the same `window' in the direction of Regions A and B. Additionally, the cooler dust, observed in the 70 $\mu$m and 160 $\mu$m {\it Spitzer} images, shows the same variation in density across the superbubble. If we assume that the \ion{H}{i} gas in the N~206 region is distributed evenly between the foreground and background of the superbubble, the amount of absorbing material in Region SB-A is at least twice that of Region A itself. Indeed, estimation of the $N_{\rm{H}}$ for [SHP2000] LMC 211 and HD 269676 from the radio flux densities at their positions yields values of $\sim 1 \times 10^{21}$ cm$^{-2}$ and $\sim 4 \times 10^{21}$ cm$^{-2}$, respectively, which is roughly consistent with the best-fit $N_{\rm{H}}$ values to their X-ray spectra. The $N_{\rm{H}}$ value of HD 269676 is slightly lower than determined from its X-ray spectrum, but this is likely due to the assumption that the \ion{H}{i} gas its direction is distributed evenly between the foreground and background of the source. We note that not all of the \ion{H}{i} gas in these considerations may be associated with the N~206 region. However, investigation of the velocity profile shows that the brightest and most significant 21 cm emission is approximately centred around a heliocentric velocity of $\sim241$~km~s$^{-1}$, which is the heliocentric velocity of the N~206 complex determined by \citet{Kim2007}. Hence, we find it likely that the variation in absorption across the face of N~206 is sufficient to obscure the X-ray emission from much of the superbubble. \\

\subsection{Properties of the X-ray emitting gas}
\par The following calculation of the physical properties of the hot gas in the superbubble follows the method outlined in \citet{Sasaki2011}. The superbubble emission is described by the harder of the two NEI components. From Equation \ref{norm} we can determine the gas density in the bubble. For the LMC metallicity $\zeta_{\rm{LMC}} = 0.5$, $n_{e} = (1.2 + 0.013\zeta_{\rm{LMC}})$ gives $n_{e} \approx 1.20 n$, with $n$ being the hydrogen density. We must also consider the value of $f_{gas}$, the filling parameter of the hot gas in the bubble and blowout. For a young stellar bubble, as is the case with N~206, the filling parameter is likely $f_{gas} \approx 1$ \citep[see][for example]{Oey2004,Sasaki2011}. Hence, we adopt this value of unity for the forthcoming calculations. Thus, Equation \ref{norm} can be written as

 \begin{equation}
\label{norm2}
norm = \dfrac{1}{10^{14}\times4\pi D^{2}} 1.2 n^{2} V 
\end{equation}

\noindent and the gas density in the bubble is given by

 \begin{equation}
\label{density}
n = 4.8 \times 10^{30} \times \sqrt{\dfrac{norm}{V}}\ [\rm{cm}^{-3}].
\end{equation}

\noindent In the case of Region A, the volume is that of a cylinder-sphere intersection with the base radius of the cylinder equal to the radius of Region A, the radius of the sphere equal to the radius of Region SB, and the cylinder edge touching the sphere edge. We determine the volume of Region A (hereafter referred to as $V_{\rm{A}}$) to be $\sim 8 \times 10^{60}\ \rm{cm}^{-3}$ and, using the NEI model normalisation of $norm_{2} = (1.3^{1.6}_{1.0})  \times 10^{-4}$, we estimate a gas density of $n_{\rm{SB}} = (1.9 \pm 0.2) \times 10^{-2}\ \rm{cm}^{-3}$.\\

\par Using this gas density and the NEI model plasma temperature $kT_{2} = 0.99 (\pm 0.36)$ keV, the pressure of the gas inside the bubble, assuming the pressure is constant throughout the entire volume, is given as

\begin{equation}
\label{pressure}
P_{\rm{SB}}/k = (n_{e} + 1.1 n) T = 2.3 n_{\rm{SB}} T_{2} = (5.1 \pm 1.9) \times 10^{5}\ \rm{cm}^{-3}\ \rm{K},
\end{equation}

\noindent With these derived values we can now estimate the thermal energy content inside the entire superbubble (Region SB), given by $E_{\rm{SB}}(t) = 3/2 P_{\rm{SB}}V_{\rm{SB}} = (3.0 \pm 1.1) \times 10^{51}\ \rm{erg}$, where $V_{\rm{SB}}$ is the volume of the superbubble sphere with $r =  4\arcmin\ (\sim 56$ pc at the LMC distance), determined from our X-ray images. In addition, the mass inside a homogeneous bubble is $M_{\rm{SB}} = 2.31n_{\rm{SB}} \times \mu \times m_{\rm{H}} \times V_{\rm{SB}} = 599 (\pm63)\ \rm{M}_{\sun}$, where $\mu = 0.61$ is the mean molecular weight of a fully ionised gas and $m_{\rm{H}}$ is the hydrogen mass. \\

\par We applied a similar treatment to the blowout region using the softer NEI component in our model with our assumed blowout region volume to derive the gas density $n_{\rm{BL}} = (1.3 \pm 0.2) \times 10^{-2}\ \rm{cm}^{-3}$, the pressure $P_{\rm{BL}}/k =  (2.4 \pm 1.2) \times 10^{5}\ \rm{cm}^{-3}\ \rm{K}$, the thermal energy content $E_{\rm{BL}}(t) = (4.7 \pm 2.4) \times 10^{50}\ \rm{erg}$, and the mass in the blowout region $M_{\rm{BL}} = 142(\pm 22)\ \rm{M}_{\sun}$.  The overall mass and thermal energy content of the superbubble and blowout region are 741$(\pm 85)$ $\rm{M}_{\sun}$ and $(3.5 \pm 1.3) \times 10^{51} \rm{erg}$.\\

\subsection{Superbubble energy budget}
The thermal energy stored in the X-ray emitting gas of the superbubble is supplied by the mechanical input of the stellar population. The pressure of the hot gas drives the expansion of the shell into the surrounding \ion{H}{i} cloud. Logically we would expect that the thermal energy stored in the hot gas and the kinetic energy of the shell  should balance the mechanical energy input of the stellar population. However, observations of X-ray superbubbles show this not to be the case. Rather, the observed energy stored in some superbubbles is much less than that injected by the stellar population \citep[see][for example]{Cooper2004,Maddox2009,Jaskot2011}. When observed superbubble properties are compared to predictions from the standard pressure-driven model of \citet{Weaver1977} or similar models, it has been found that the predicted dynamical evolution of the superbubble for a given stellar input proceeds at a slower rate than is observed. This is known as the growth rate discrepancy.  It is clear that much of the energy injected into the superbubbles by the stellar populations is `missing' and several mechanisms, from superbubble blowouts to non-thermal X-ray production, have been proposed to explain this problem. Detailed discussions on this subject can be found in \citet{Cooper2004} and \citet{Jaskot2011}. We now examine N~206 with regard to the growth rate discrepancy, determining the kinetic energy stored in the shell and, as the ionisation front may be trapped in the shell, the surrounding \ion{H}{i} gas, as well as the energy supplied by the stellar population.

\subsubsection{Energy stored in the H$\alpha$ shell}
To determine the kinetic energy stored in the H$\alpha$ shell, we must first determine its mass. This can be derived from the H$\alpha$ luminosity of the shell, $L_{\rm{H}\alpha}$. Using scans of the Curtis Schmidt plates of \citet{Kennicutt1986},  \citet{Dunne2001} determined the shell $L_{\rm{H}\alpha}$ to be $\sim 9 \times 10^{36}\ \rm{erg\ s}^{-1}$. For ionised gas at temperatures of $\sim10^{4}$ K, $L_{\rm{H}\alpha} = 3.56 \times 10^{-25}n_{e}n_{p}Vf_{shell}$, where $n_{e}$ and $n_{p}$ are the number densities of electrons and protons, respectively, $V$ is the volume and $f_{shell}$ is the filling parameter. Helium is likely singly ionised, so  $n_{e} = 1.1n_{p}$. Substituting this into the equation above with the known value of $V$, we obtain $n_{e} = 1.02f_{shell}^{-1/2}$ cm$^{-3}$. The total mass in the superbubble shell is then given by $M_{sh} = 1.27n_{e}Vf_{shell} m_{\rm{H}} \approx 3 \times 10^{4}\ \rm{M_{\sun}}\ f_{shell}^{1/2}$. The filling parameter $f_{shell}$ is determined from the thickness of the H$\alpha$ shell, which in the case of our superbubble is $\sim10-15$\% of the superbubble radius. This corresponds to a filling parameter $f_{shell} = 0.35(\pm 0.05)$, yielding a total mass of $(1.8 \pm 0.2) \times 10^{4}\ \rm{M_{\sun}}$. We know from \citet{Dunne2001} that the expansion velocity of the H$\alpha$ shell is $\sim30$ km s$^{-1}$, thus we derive a kinetic energy $E_{\rm{k}} = \frac{1}{2}Mv_{\rm{exp}}^{2} \approx (1.6\ \pm 0.2) \times 10^{50} \rm{erg}$. 

\subsubsection{Energy stored in \ion{H}{i} gas}
\label{HIgas}
The mass and energy input of the stellar populations in the superbubble drive the shell into the surrounding \ion{H}{i} gas. If the ionisation front of the stellar population becomes trapped in the shell, the outer layer remains cold \ion{H}{i} gas. The mass of this swept-up \ion{H}{i} gas can be determined from the 21 cm line emission line data from the ATCA-Parkes survey of the Magellanic Clouds, used to produce Fig. \ref{HI_image}. We determine the total line flux density from the region of the superbubble and blowout (shown as the blue region in Fig. \ref{HI_image}) to be $\sim238$ Jy. This corresponds to a mass of $\sim 1 \times 10^{5}\ \rm{M_{\sun}}$ at the LMC distance of 48 kpc. We assumed that the expansion velocity of the swept up \ion{H}{i} gas is the same as that of the H$\alpha$ shell, namely, $\sim30$ km s$^{-1}$. This implies the kinetic energy stored in the \ion{H}{i} gas is $\sim 1 \times 10^{51}$ erg. However, this value should be considered an upper limit as not all of the \ion{H}{i} gas in the region may be associated with the superbubble.

\subsubsection{Stellar content and input}
\label{content}
To date, the high-mass stellar population of the N~206 \ion{H}{ii} region has not been the subject of a detailed spectroscopic analysis. However, the photometric  study of high-mass stars in the Magellanic Clouds of \citet{Massey1995} provides us with the spectral types of the massive stars in N~206. Additionally, more recent analyses of specific source classes in the LMC \citep{Brey1999,Robert2003} supplement and/or update the spectral types of \citet{Massey1995}. We used the SIMBAD database to determine the most recent information available for the OB stars in the N~206 region and adopted the determined spectral types. Association membership was assigned based on the location of the stars with respect to the association dimensions for LH 66, LH 69, and NGC 2018, as determined by \citet{Bica1999}. Since we only have limited data on these sources, it is difficult to say whether they indeed belong to the stellar associations or just appear to be due to projection effects.  Nevertheless, we list the high-mass stars with known spectral types with their likely host association in Table \ref{sources}.\\

\begin{table*}[htdp]
\caption{High-mass stellar content of superbubble. $v_{\infty}$ and $\dot{M}$ are the estimated terminal wind velocity and mass loss rates, respectively (see text)}.
\begin{center}
\begin{footnotesize}
\label{sources}
\begin{tabular}{lllllcc}
\hline
RA & DEC & Object Name  &  Association & Spectral Type &  $v_{\infty}$ & $\dot{M}$ \\
       &          &                           &                        &                           & (km s$^{-1}$)                            & ($\rm{M}_{\sun}$ yr$^{-1}$)  \\
\hline
\hline
05:29:52	&	-71:04:07	&	[MLD95] LMC 1-548 	&	LH 66	&	B1.5V 	&	1790	&	$5.4 \times 10^{-8}$	\\
05:29:52	&	-71:04:33	&	2MASS J05295152-7104325	&	LH 66	&	O8V 	&	2094	&	$4.3 \times 10^{-7}$	\\
05:30:11	&	-71:04:04	&	2MASS J05301149-7104034	&	LH 66	&	B1III 	&	1598	&	$2.8 \times 10^{-7}$	\\
05:30:11	&	-71:04:47	&	2MASS J05301122-7104464 	&	LH 66	&	O9V 	&	2039	&	$2.9 \times 10^{-7}$	\\
05:30:12	&	-71:06:00	&	2MASS J05301167-7105596	&	LH 66	&	B0.5V 	&	1872	&	$1.9 \times 10^{-7}$	\\
05:30:19	&	-71:06:07	&	2MASS J05301948-7106071	&	LH 66	&	B0V 	&	1942	&	$3.1 \times 10^{-7}$	\\
05:30:23	&	-71:06:19	&	2MASS J05302268-7106194	&	LH 66	&	O8V 	&	2066	&	$4.4 \times 10^{-7}$	\\
05:30:25	&	-71:03:45	&	2MASS J05302469-7103445	&	LH 69	&	O7V 	&	2081	&	$7.0 \times 10^{-7}$	\\
05:30:31	&	-71:02:32	&	2MASS J05303066-7102315	&	LH 69	&	B0.5V 	&	1872	&	$1.9 \times 10^{-7}$	\\
05:30:35	&	-71:01:57	&	2MASS J05303520-7101570	&	LH 69	&	B1Ve 	&	1829	&	$1.0 \times 10^{-7}$	\\
05:30:37	&	-71:01:43	&	[MLD95] LMC 1-715 	&	LH 69	&	B0.5V 	&	1872	&	$1.9 \times 10^{-7}$	\\
05:30:37	&	-71:04:16	&	2MASS J05303660-7104158	&	LH 69	&	O9V 	&	2039	&	$3.0 \times 10^{-7}$	\\
05:30:39	&	-71:01:48	&	HD 37248 	&	LH 69	&	WC4+O9 	&	2400	&	$3.5 \times 10^{-5}$	\\
05:30:45	&	-71:01:45	&	[MLD95] LMC 1-713 	&	LH 69	&	B1III 	&	1598	&	$2.8 \times 10^{-7}$	\\
05:30:47	&	-71:03:16	&	2MASS J05304707-7103156	&	LH 69	&	B0Ia 	&	1369	&	$2.2 \times 10^{-6}$	\\
05:30:48	&	-71:04:02	&	HD 269660	&	LH 69	&	B1Iab 	&	1189	&	$1.5 \times 10^{-6}$	\\
05:30:51	&	-71:02:48	&	2MASS J05305063-7102480 	&	LH 69	&	O8V 	&	2034	&	$5.2 \times 10^{-7}$	\\
05:30:54	&	-71:05:11	&	2MASS J05305448-7105105	&	LH 69 	&	B1II	&	1598	&	$2.8 \times 10^{-7}$	\\
05:30:58	&	-71:01:32	&	2MASS J05305766-7101315	&	LH 69	&	B1III 	&	1598	&	$2.8 \times 10^{-7}$	\\
05:30:58	&	-71:03:34	&	2MASS J05305760-7103337	&	LH 69	&	O8V... 	&	2066	&	$4.4 \times 10^{-7}$	\\
05:31:10	&	-71:03:52	&	2MASS J05310966-7103515	&	LH 69	&	B1III 	&	1598	&	$2.8 \times 10^{-7}$	\\
05:31:11	&	-71:01:43	&	2MASS J05311070-7101433	&	LH 69	&	O8III... 	&	1788	&	$1.3 \times 10^{-6}$	\\
05:31:12	&	-71:02:58	&	[MLD95] LMC 1-485 	&	LH 69	&	B0Ve 	&	1942	&	$3.1 \times 10^{-7}$	\\
05:31:12	&	-71:04:10	&	[MLD95] LMC 1-550	&	NGC 2018	&	O5V 	&	2095	&	$1.8 \times 10^{-6}$	\\
05:31:14	&	-71:03:51	&	[MLD95] LMC 1-576	&	NGC 2018	&	O7.5V 	&	2074	&	$5.6 \times 10^{-7}$	\\
05:31:16	&	-71:04:10	&	HD 269676 	&	NGC 2018	&	O4-5III(f) 	&	1802	&	$3.2 \times 10^{-6}$	\\
05:31:16	&	-71:03:43	&	[MLD95] LMC 1-552	&	NGC 2018	&	O4Iab:... 	&	1435	&	$8.0 \times 10^{-6}$	\\
05:31:19	&	-71:04:39	&	2MASS J05311887-7104389	&	NGC 2018	&	O7Ve 	&	2081	&	$7.0 \times 10^{-7}$	\\
05:31:22	&	-71:04:06	&	2MASS J05312153-7104057	&	NGC 2018	&	O8V 	&	2066	&	$4.4 \times 10^{-7}$	\\
05:31:24	&	-71:04:13	&	[MLD95] LMC 1-546 	&	NGC 2018	&	O6Ve 	&	2094	&	$4.2 \times 10^{-6}$	\\

\hline
\end{tabular}
\end{footnotesize}
\end{center}
\end{table*}%

\noindent To determine the mass and energy input of the massive stars, we followed a similar method to that of \citet{Sasaki2011}. We assigned effective temperatures and luminosities from \citet{Martins2005} for the O stars and \citet{SK1982} for the B stars, and used the mass-luminosity relation of Vitrichenko et al. (2007) to obtain the mass ($M$) of each star:

\begin{equation}
\label{lum}
L = 19(M/M_{\sun})^{2.76}L_{\sun}.
\end{equation}

\noindent The radii of the stars are obtained from

\begin{equation}
\label{radius}
R = \sqrt{\dfrac{L}{4\pi \sigma T^{4}_{\rm{eff}}}},
\end{equation}

\noindent where $T_{\rm{eff}}$ is the effective temperature of the star and $\sigma$ is the Stefan-Boltzmann constant.  The wind velocity was then calculated using the theory of radiation-driven winds \citep{Castor1975},

\begin{equation}
\label{winds}
v_{\infty} = av_{\rm{esc}} = a\left[\dfrac{2GM}{R} \times (1-L/L_{\rm{edd}})\right]^{0.5},
\end{equation}

\noindent where $a = 2.6 (\pm 0.2)$ \citep{Lamers1995}, $v_{\rm{esc}}$ is the photospheric escape velocity and $L_{\rm{edd}}$ is the Eddington luminosity given by $L_{\rm{edd}} = 4\pi G \times M \times m_{p} \times c/\sigma_{T}$ with $m_{p}$ being the proton mass and $\sigma_{T}$ the Thompson cross section for the electron. The mass loss is determined from the single scattering limit

\begin{equation}
\label{mloss}
\dot{M} = \dfrac{L}{v_{\infty} \times c}.
\end{equation}

\noindent We corrected the wind velocities and mass loss rates for LMC metallicity following \citet{Leitherer1992} and \citet{Vink2001}, respectively. The estimates for wind velocities and mass loss rates, which are shown in Table \ref{sources}, are in general agreement with the wind properties of LMC stars in the literature \citep[see][for example]{Mokiem2007}. \\

\par
\begin{figure*}
\begin{center}
\resizebox{\hsize}{!}{\includegraphics[trim= 1cm 1cm 1cm 1cm, clip=true]{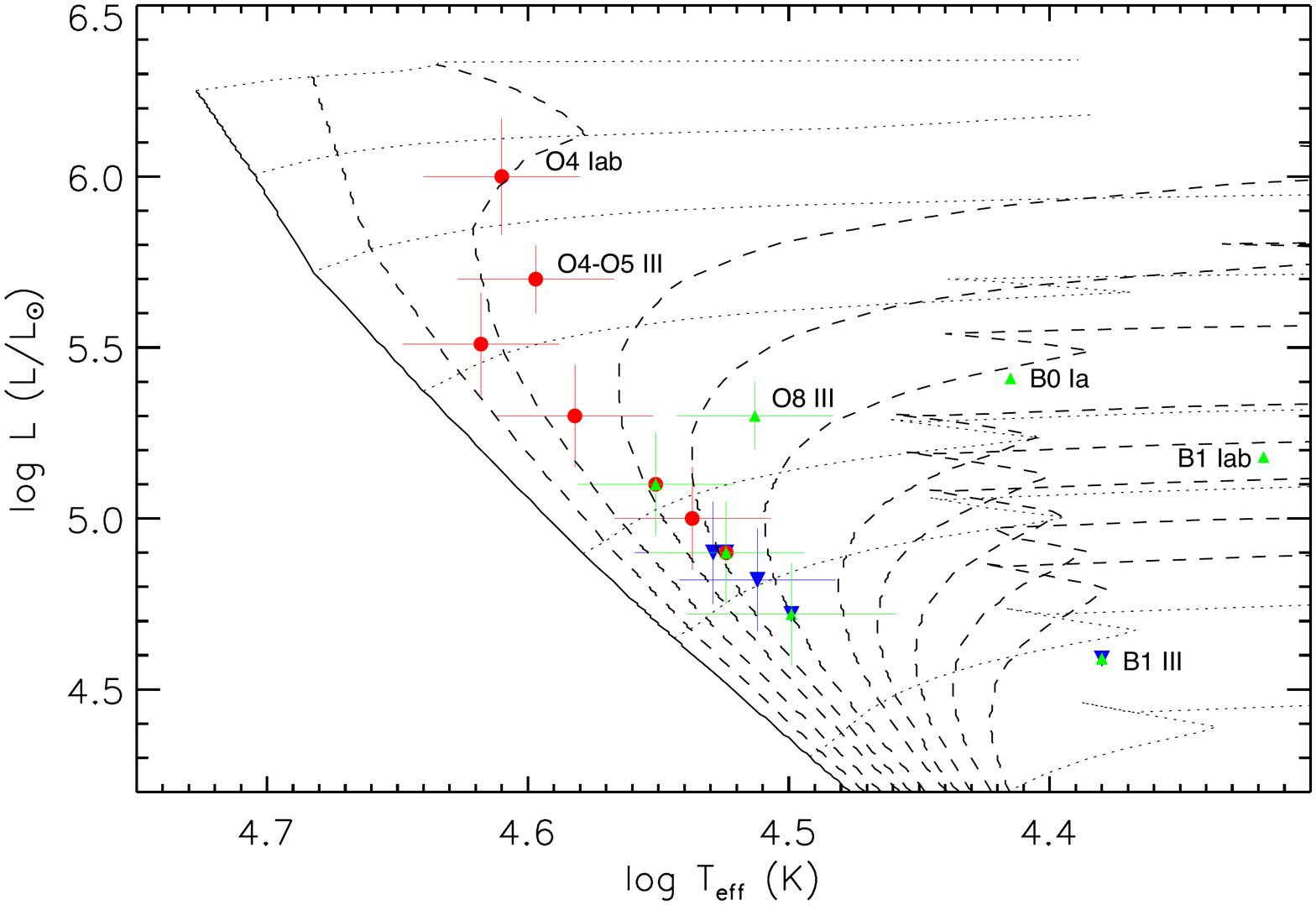}}
\caption{HR diagram showing the stellar populations of LH 66, LH 69, and NGC 2018 compared to the evolutionary tracks and isochrones of \citet{Schaller1992}. LH 66 sources are indicated by the blue inverted triangles, LH 69 sources by the green triangles, and NGC 2018 sources by the red circles. The solid black line is the ZAMS isochrone with the dashed lines showing isochrones from 1-10 Myr in increments of 1 Myr. The dotted lines show the evolutionary tracks of stars with masses of (from top to bottom) 120 M$_{\sun}$, 85 M$_{\sun}$, 60 M$_{\sun}$, 40 M$_{\sun}$, 25 M$_{\sun}$, 20 M$_{\sun}$, 15 M$_{\sun}$, and 12 M$_{\sun}$. For clarity, we only show part of the evolutionary tracks of the most massive stars. We also omit the main-sequence B stars because they are unimportant in this age determination. The spectral types of the post-main sequence stars are also labelled to aid in the discussion. Uncertainties are unavailable for the B stars. }
\label{tracks}
\end{center}
\end{figure*}

 To determine the total mass and energy injected into the superbubble by the stellar populations we must first determine their ages. However, because no high-quality optical data are available, robust age determination methods such as theoretical isochrone fitting are beyond our grasp. One possibility is to assume the age of the superbubble as the age of the stellar populations. \citet{Dunne2001} gave the expansion velocity of the superbubble H$\alpha$ shell as $\sim 30$ km s$^{-1}$. Because the superbubble has expanded to a diameter of $\sim$ 112 pc, this suggests an age of $\sim$ 2 Myr. However, the expansion velocity of $\sim 30$ km s$^{-1}$ falls in the high-velocity superbubble regime \citep{Oey1996}, meaning the N~206 superbubble must have undergone induced acceleration at some point in its history, most likely due to SNRs impacting the outer shell. Thus, the superbubble is likely older than 2 Myr and, as such, is not a good estimate for the age of the stellar populations. \citet{Gorjian2004} used previous radio and optical photometric measurements of \citet{Kim1999} and \citet{Bica1996}, respectively, to constrain the age of the stellar populations to 2-10 Myr. However, this wide range of possible ages does not allow the total mass and energy input of the stellar population to be sufficiently constrained. Instead we took the effective temperatures and luminosities used above to produce HR diagrams for each association to compare them with theoretical isochrones. While this is not a perfect solution, it at least allows us to obtain better age estimates than are currently available. Since we assigned effective temperatures and luminosities at Galactic metallicity\footnote{We corrected the results of our mass loss and terminal velocity calculations for metallicity effects and not the input effective temperatures and luminosities, which are representative of solar metallicity.}, we used the theoretical isochrones of \cite{Schaller1992} at $z = 0.020$ to gauge the likely ages of the stellar associations. The HR diagram is shown in Fig. \ref{tracks}. \\

\par The age of the NGC 2018 population is well described by the 2 Myr isochrone. Owing to slight scatter in the points and the magnitude of the error bars, we assign an age range of $2(\pm1)$ Myr for NGC 2018. With regard to LH 66 and LH 69, the presence of the B1 III stars, which indicate an age $>10$ Myr, is somewhat at odds with the main-sequence populations, which are indicative of much younger ages. Because of this, we suggest that the B1 III sources are not associated with LH 66 or LH 69. This is a reasonable conclusion because the wider N~206 \ion{H}{ii} region hosts stellar populations outside of the superbubble we discuss in this paper. Due to dynamical ejection or supernova kick processes, a star with a modest ejection/kick velocity of $\sim 30$ km s$^{-1}$ can cross the diameter of the superbubble ($\sim 112$ pc) in  $<5$ Myr, well within the lifetime of early B-type stars. The remaining LH 69 stars show a reasonably large spread in age but the location of the O8 III and B0 Ia stars suggest an age of $\sim 5$ Myr. The B1 Iab star is older than this while the top of the main sequence suggests an age younger than 5 Myr. Therefore, we set a cautious age range of $5(\pm3)$ Myr. Unfortunately, the LH 66 population has no post-main sequence objects. We assigned its age based on its upper main sequence to be $4(\pm2)$ Myr. The age determinations for each of the stellar associations are consistent with the 2-10 Myr age constraint provided by \citet{Gorjian2004}. However, our age estimations consider only single-star evolution and not close binary evolution effects. \citet{Van1998} showed that the majority of massive stars are formed in interacting close binary systems and that mass transfer effects in these binaries can lead to incorrect age determinations for massive stellar clusters older than $\sim4$~Myr through to the process of `starburst rejuvenation'. This leads to a possible underestimation of cluster age of up to a few Myr if close binary evolution is not considered. For the N~206 associations with ages $\gtrsim4$~Myr, namely LH~66 and LH~69, the determined age ranges are large due to the limitations of the available data, but may sufficiently account for this underestimation of cluster age. This question can only be resolved with higher quality optical/IR data and the application of binary evolutionary tracks.\\

\par Using these determined ages, mass loss rates, and wind velocities, we estimate that the O and early-B stars in the superbubble have supplied a mass of $86(\pm52)\ \rm{M}_{\sun}$ and a total energy of $(2.5 \pm 1.5) \times 10^{51}\ \rm{erg}$. \\

\par The single Wolf-Rayet (WR) star in the superbubble HD~37248, which is of carbon class (WC) and a member of LH 69, also supplies a large amount of mass and energy. According to the theoretical evolutionary tracks at LMC metallicity of \citet{Schaerer1993}, only a star more massive than 60 M$_{\sun}$ will enter the WC phase. Given the large error on the age estimate for LH 69, namely $5(\pm3)$ Myr, it is impossible to constrain the initial mass of the WC star further. We adopted mass loss rates and velocities from \citet{Leitherer1997} and corrected the mass loss rates for the metallicity of the LMC according to \citet{Crowther2007} (listed in Table \ref{sources}), the results of which are in rough agreement with those of other WC class stars in the LMC from \citet{Grafener1998}. We calculated the total mass and energy input over the WR lifetimes \citep[from][]{Schaerer1993} for stars of mass $>60\ \rm{M}_{\sun}$, assuming they are half way through their WC lifetime. Additionally, we calculated the mass and energy input of the star before it entered the WR phase, which we estimated using the methods outlined above and the pre-WR lifetimes of stars $>60\ \rm{M}_{\sun}$ from \citet{Schaerer1993}. We find that over its lifetime the WC star has injected $20(\pm3) \rm{M}_{\sun}$ and $(1.1 \pm 0.2) \times 10^{51} \rm{erg}$ into the superbubble. The mass loss rates and wind velocity estimates are determined for the evolution of a single WC star. Because HD~37248 is in a binary system, its age estimate and wind parameters are more uncertain. For example, \citet{Sander2012} showed that stellar wind parameters of binary WC stars are different from single WC stars of the same spectral type (see their Fig. 6). However, without high-quality optical spectra we cannot further constrain the wind parameters of this object.\\

\noindent We must also estimate the contribution of previous supernovae (SNe). We assumed that the stellar populations formed from the same parental molecular cloud and are described by a standard Salpeter IMF \citep{Salpeter1955}. We find from Fig. \ref{tracks} that there are some 16 stars in the 20-40 M$_{\sun}$ bins and 4 stars with masses $> 40\ \rm{M}_{\sun}$, including the WR star. Hence, we determined that $3(\pm2)$ SN have already occurred in the bubble. Taking an SN explosion energy of $10^{51}$ erg, these SNae have injected $\sim (3 \pm 2) \times 10^{51}$ erg into the superbubble and, assuming an average initial mass of $60\ \rm{M}_{\sun}$, supplied some $60-300\ \rm{M}_{\sun}$ of material through their ejecta.\\

\noindent Thus, in total the stellar population has injected $286(\pm175) \rm{M}_{\sun}$ and  $(6.6 \pm 3.7) \times 10^{51}\ \rm{erg}$ into the bubble. \\

\subsubsection{Observed/expected energy and mass comparison}
The observed and expected energies and masses in the N~206 superbubble are summarised in Table \ref{budget}. In total, the overall energy input by the stellar populations is $(6.6 \pm 3.7) \times 10^{51}\ \rm{erg}$ while the observed energy stored in the superbubble is $\lesssim (4.7 \pm 1.3) \times 10^{51}\ \rm{erg}$. Hence, no definitive conclusions can be drawn regarding the energy budget in the N~206 superbubble. Because of the limited data on the stellar populations, the uncertainty on the age estimates of the stellar associations is substantial, making the uncertainty on the stellar energy input quite large. Only when high-quality optical/NIR data are available for N~206 will the stellar energy input be sufficiently constrained to say for sure.\\

\par The overall mass contained in the superbubble and blowout region is 741 $(\pm 85)$ $\rm{M}_{\sun}$. We expect the stellar populations to have contributed $286(\pm175) \rm{M}_{\sun}$. Even with the large uncertainties on the mass from the stellar populations arising from the large uncertainty in ages, the observed mass is much higher. This higher-than-expected observed mass is likely due to mass loading by evaporation of entrained interstellar clouds and/or turbulent mixing of material from the cold shell \citep{Weaver1977,Silich1996}. 

\begin{table}
\caption{Observed/expected energies and masses in the N~206 superbubble. Observed values are calculated assuming uniform gas density and a filling factor of unity.}
\begin{center}
\begin{footnotesize}
\label{budget}
\begin{tabular}{lcc}
\hline
\hline
 & Energy & Mass \\
	 & ($10^{51}$ erg s$^{-1}$) & (M$_{\sun}$)  \\
\hline
\multicolumn{3}{c}{{\tiny Observed}} \\
\hline
X-ray emitting gas  & $3.5 (\pm 1.3)$ & $741 (\pm 85)$\\
Kinetic energy of H$\alpha$ shell & $0.16 (\pm 0.02)$ & - \\
Kinetic energy of surrounding \ion{H}{i} gas  & $\sim 1$ & - \\
Total observed & $4.7 (\pm 1.3)$ & $741 (\pm 85)$\\
\hline
\multicolumn{3}{c}{{\tiny Expected Input}} \\
\hline
Stellar wind input & $3.6 (\pm 1.7)$ & $106 (\pm 55)$\\
Supernovae input & $3 (\pm 2)$ & $180 (\pm 120)$ \\
Total input & $6.6 (\pm 3.7)$ & $286 (\pm 175)$\\
\hline
\end{tabular}
\end{footnotesize}
\end{center}
\end{table}%

\begin{figure*}
\begin{center}
\resizebox{\hsize}{!}{\includegraphics{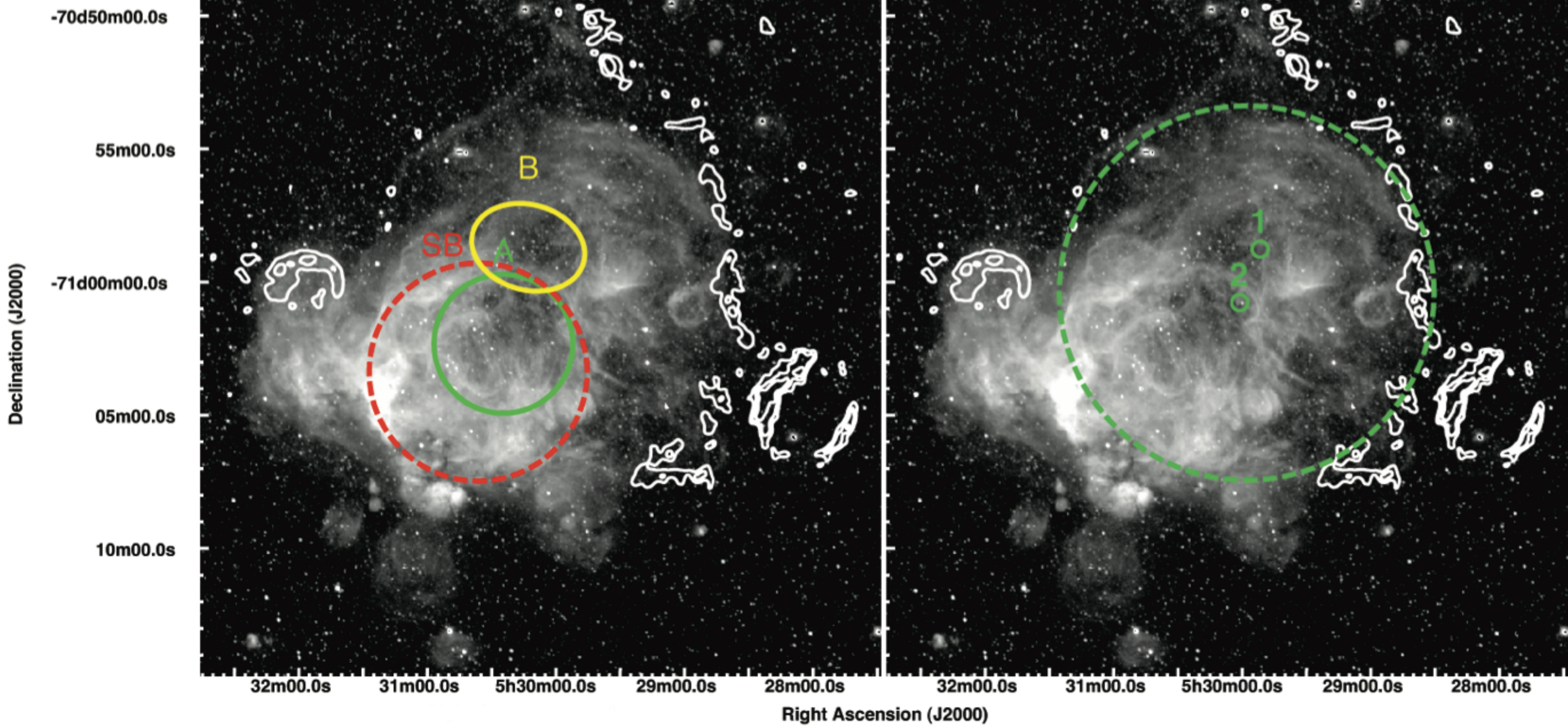}}
\caption{ MCELS H$\alpha$ image of the N~206 \ion{H}{ii} region. Left: N206 superbubble and blowout regions shown with the white contours indicating regions with [\ion{S}{ii}]/H$\alpha > 0.67$. Right: Known HMXB XMMU J052952.9-705850 and candidate HMXB USNO-B1.0 0189-00079195 indicated by the small circles labelled 1 and 2, respectively. The large dashed green circle marks the apparent shell of the candidate old SNR. Again the white contours indicate regions with [\ion{S}{ii}]/H$\alpha > 0.67$. Both SNR B0532-71.0 and the candidate SNR [HP99]~1234 are evident in the east and west, respectively.}
\label{snr}
\end{center}
\end{figure*}

\subsubsection{Evidence for an old SNR in N~206?}
During our analysis of the N~206 superbubble, we attempted to identify signatures of SNR impacts of the superbubble shell walls using the well-known optical identification method of the [\ion{S}{ii}]/H$\alpha$ ratio, with a value of [\ion{S}{ii}]/H$\alpha \gtrsim 0.4$ being indicative of a shock-heated nebula \citep{Matt1973,Raymond1979}. However, as found by \citet[][and references therein]{Fesen1985}, if an SNR is located in an \ion{H}{ii} region, a higher [\ion{S}{ii}]/H$\alpha \gtrsim 0.67$ is required to unambiguously attribute the optical emission to an SNR, particularly in the absence of additional SNR identification criteria such as X-ray or radio emission. To assess the [\ion{S}{ii}]/H$\alpha$ ratio in the N~206 region, we used the MCELS data to create continuum-subtracted [\ion{S}{ii}] and H$\alpha$ images, and produced an [\ion{S}{ii}]/H$\alpha$ ratio image. Contours indicating the regions of N~206 with [\ion{S}{ii}]/H$\alpha > 0.67$ were derived from this image and are plotted on the MCELS H$\alpha$ in Fig. \ref{snr}. We found no evidence for SNRs in the X-ray superbubble. Interestingly, however, we find a much larger, approximately semi-circular shell of shock heated material in the wider N~206 complex (highlighted as the dashed green circle in Fig. \ref{snr}), as well as SNR B0532-71.0 in the east and the candidate SNR [HP99]~1234 \citep{Haberl1999} in the west. A study of the candidate SNR [HP99]~1234 will be published in a separate paper \citep{Kavanagh2012b}. Intriguingly, the known HMXB and candidate HMXB in the complex are both located close to the apparent centre of the large optical shell (see Fig. \ref{snr}). We determine the diameter of the shell to be $\sim 195$ pc. This is certainly large for an SNR, considering the largest confirmed SNRs in the LMC, such as SNR~0450$-$70.9 \citep{Matt1985,Williams2004,Cajko2009}, LMC~SNR~J0550$-$6823 \citep{Davies1976,Filipovic1998,Bozzetto2012} and SNR0506-6542 \citep{Klimek2010}, are significantly smaller. Larger SNR diameters can be achieved if the ejecta expand into a low-density medium as is the case for the galactic Monogem Ring. \citet{Plucinsky1996} determined the diameter of the Monogem Ring to be $\sim 133$ pc and showed that this SNR has evolved in a region of the Galaxy with abnormally low density. These two examples point to a possible explanation for the large diameter of the optical shell. We suggest that, in the case of the large optical shell, the ejecta have been expanding into a wind-blown bubble, likely a superbubble, and are only now impacting and shock-heating the superbubble shell walls, producing the [\ion{S}{ii}] enhancement we observe. As discussed by \citet{Chu1997}, this is an entirely plausible scenario. An SNR deep in a superbubble will not be detectable until the ejecta shock dense material, such as the shell walls or cloudlets in the superbubble interior. It is not unreasonable to assume the presence of a second older superbubble in the N~206 complex. As mentioned in our analysis, the wider complex hosts other OB associations. The association BSDL1927 \citep{Bica1999} is located outside of the superbubble discussed above, but inside the large optical shell. Using the SIMBAD database we determine that this association contains a small number of later O stars and a WR star. Indeed, the known HMXB, XMMU J052952.9-705850, is located $\sim2'$ east of this association. Unfortunately, the {\it XMM-Newton} observations do not cover the west of the large optical shell where the shocks are located (see Fig. \ref{snr}) so it is uncertain whether the shell is X-ray bright. Only high-quality X-ray and radio data will resolve the physical nature of this object.

\section{Summary}
Using {\it XMM-Newton} observations, we have detected significant soft diffuse X-ray emission in the N~206 \ion{H}{ii} region. We performed a morphological comparison of the X-ray emission to H$\alpha$ emission in N~206 utilising MCELS data and found that the brightest X-ray emission is confined by an H$\alpha$ structure that extends as an approximately circular shell. This points to a superbubble origin for the detected emission where a hot shocked X-ray emitting gas is confined by the cooler swept-up material, confirming the results of \citet{Dunne2001}. In addition, faint diffuse X-ray emission was detected extending beyond the H$\alpha$ shell, which we attribute to a blowout region where hot gas from the superbubble is escaping into the surrounding ISM. We performed spectral analyses on the superbubble and blowout emission, finding both to be thermal in origin, consistent with a hot shocked gas. We also used the MCELS data to identify a large semi-circular shell of shock-heated material with the optical emission line properties of an SNR. Because of the large diameter ($\sim 195$ pc), we suggest it is likely an old SNR that has expanded into a wind-blown bubble and is only now interacting with the bubble walls to produce the optical emission we observe. \\

\par Using the results of the N~206 superbubble spectral analysis and assumed volumes, we calculated the physical properties of the hot gas in the bubble and blowout. We determined a temperature of $kT = 0.99 (\pm 0.36)$ keV, a density of $n_{\rm{SB}} = (1.9 \pm 0.2) \times 10^{-2}\ \rm{cm}^{-3}$ and a pressure of $P_{\rm{SB}}/k = (5.1 \pm 1.9) \times 10^{5}\ \rm{cm}^{-3}\ \rm{K}$ for the hot gas in the superbubble and corresponding values of  $kT = 0.70 (\pm 0.31)$ keV, $n_{\rm{BL}} = (1.3 \pm 0.2) \times 10^{-2}\ \rm{cm}^{-3}$ and $P_{\rm{BL}}/k =  (2.4 \pm 1.2) \times 10^{5}\ \rm{cm}^{-3}\ \rm{K}$ for the blowout. Using these values, we determined the overall mass and thermal energy content of the superbubble and blowout region to be 741$(\pm 85)$ $\rm{M}_{\sun}$ and $(3.5 \pm 1.3) \times 10^{51} \rm{erg}$. \\

\par To determine if the N~206 superbubble exhibits the same growth rate discrepancy as observed in other superbubbles, we compared the observed energy stored in the superbubble to the expected energy supplied by its stellar population. We combined the thermal energy stored in the hot gas with the kinetic energy of the expanding H$\alpha$ shell and surrounding \ion{H}{i} gas to determine a total observed energy of $\lesssim (4.7 \pm 1.3) \times 10^{51}\ \rm{erg}$. We used the available information of the N~206 stellar population to determine the expected energy from stellar winds and past SNRs of $(6.6 \pm 3.7) \times 10^{51}\ \rm{erg}$. Due to the poorly constrained stellar population input we cannot definitively confirm a growth rate discrepancy for the N~206 superbubble. Only detailed optical/NIR analysis of the stellar population in the superbubble will constrain the input sufficiently to provide a conclusive answer.

\begin{acknowledgements}
We wish to thank the anonymous referee for very constructive suggestions to improve the paper. This research is based on data obtained with {\it XMM-Newton}, an ESA science mission with instruments and contributions directly funded by ESA Member States and NASA. P.K. is funded through the BMWI/DLR grant FKZ 50 OR 1009. M. S. acknowledges support by the Deutsche Forschungsgemeinschaft through the Emmy Noether Research Grant SA 2131/1. The MCELS data are kindly provided by R.C. Smith, P.F. Winkler, and S.D. Points. The MCELS project has been supported in part by NSF grants AST-9540747 and AST-0307613, and through the generous support of the Dean B. McLaughlin Fund at the University of Michigan. The National Optical Astronomy Observatory is operated by the Association of Universities for Research in Astronomy Inc. (AURA), under a cooperative agreement with the National Science Foundation. This research has made use of the SIMBAD database operated at CDS, Strasbourg, France.

\end{acknowledgements}

\bibliographystyle{aa}
\bibliography{refs.bib}

\end{document}